\documentclass{article}
\usepackage{amsmath,amssymb,mathrsfs,hyperref}

\newcommand{\ket}[1]{|#1\rangle}
\newcommand{\bra}[1]{\langle#1|}
\newcommand{\bracket}[2]{\langle#1|#2\rangle}

\newcommand{\wick}[1]{\pmb{:}{\!}{\!}#1{\!}{\!}\pmb{:}}
\newcommand{\qwick}[1]{\pmb{:}{\!}{\!}#1{\!}{\!}\pmb{:}{\!\!}_{\scriptscriptstyle{Q}}}

\title{Aspects of Quantum Field Theory\\ on\\ Quantum Spacetime}

\author{Gherardo Piacitelli}

\date{{\small SISSA}\\{\small
        Via Bonomea 265, 34136 - Trieste (Italy)}\\
       {\small e-mail: {\tt gherardo@piacitelli.org}}}

\begin{document}
\maketitle

\begin{abstract}
We provide a minimal, self-contained introduction to the 
covariant DFR flat 
quantum spacetime, and to some partial results for the corresponding quantum 
field theory. Explicit equations are given in the Dirac notation.
\end{abstract}

\tableofcontents

\section{Introduction}

At the time of the beautiful conference in Corfu\footnote{
Corfu Summer Institute on Elementary Particles and Physics - Workshop on Non Commutative Field Theory and Gravity, September 8-12, 2010, Corfu Greece.}, 
an extensive review 
on these and related topics had already been completed, 
which now is published as \cite{Piacitelli:2010bn}.
I will then take the opportunity of the present conference proceedings to 
complement that review with an explicit description of some applications
to quantum field theory 
\cite{Doplicher:1994tu,Bahns:2003vb,Bahns:2004fc,Piacitelli:2004rm} on DFR 
quantum spacetime \cite{Doplicher:1994tu}. 
The presentation is however self contained and 
addresses a possibly different audience. The second chapter partially 
overlaps with \cite{Piacitelli:2010bn}, but contains more explicit equations. The third 
chapter contains a much more detailed description of results about quantum 
field theory than \cite{Piacitelli:2010bn}. For the 
sake of dissemination, I will give explicit formulas in the Dirac 
notation, still favoured by many physicists. The reader interested in 
mathematical rigour is referred to the original papers.

In this first introductory chapter, 
we will very briefly recall some basics about local 
quantum field theory
on the flat Minkowski spacetime in physical (i.e.\ 4) dimensions
(section \ref{subsec:lqft}). 
Then we will discuss motivations for introducing non commutative
coordinates, and their interpretation: in particular we will stress that
they are NOT observables (section \ref{subsec:why}).

As for the comparison with other approaches to covariance, the interested 
reader may find a detailed discussion in \cite{Piacitelli:2010bn}.

\subsection{A minimal account of local quantum field theory}
\label{subsec:lqft}
Relativistic quantum field theory results from merging the quantum theory
of observables with the principle of locality: it is formulated in terms
of operator fields 
\(A(x)\), which can be regarded (up to idealisations) as sets of pointwise 
localised observables, labeled by the event \(x\) at which they are localised;
equivalently as more or less generalised ``functions'' of \(\mathbb R^4\).

Einstein causality amounts to require the commutativity of any two observable
fields \(A(\cdot),B(\cdot)\) at spacelike distances, namely \([A(x+a),B(x)]=0\)
for every \(a\) spacelike.

There is a representation \(U(\varLambda,a)\) of
the Poincar\'e group. Then a (scalar) operator field \(A(\cdot)\) 
is said covariant if 
\(U(\varLambda,a)A(x)U(\varLambda,a)^{-1}=A(\varLambda x+a)\).
Covariance is required so that every observable which is at rest in given frame
can be described by any other equivalent observer (passive point of view); and 
also so that the measuring device can be displaced, rotated and boosted 
so to be brought at rest with respect to any other equivalent observer (active point of view). A covariant field, then, essentially describe a single device
in all possible Lorentz frames.

For general reasons (spin-statistics theorem), possibly unobservable fields 
also must be allowed 
for, which either commute or anticommute at spacelike distances; for these and 
related reasons, covariance has to be 
generalised to multiplets of such fields. Here however we will confine
ourselves with a theory generated by a single covariant Bosonic field.

The pairwise 
commuting generators \(P^\mu\) of translations, defined by \(U(I,a)=e^{iaP}\), 
fulfil the spectrum condition: \(P^0\geqslant 0\) and 
\({P^0}^2-|\vec P|^2\geqslant 0\). 
In particular \(H=P^0\) is the generator of time translations=time evolution, 
and is called the Hamiltonian (with respect to the given Lorentz frame).

The vacuum state \(\ket\emptyset\) is defined as the unique translation invariant 
state (if symmetries are not spontaneously broken), 
thus fulfilling \(P^\mu\ket\emptyset=0\). If the theory is defined by a single
field \(A(x)\), any state 
\(|\Psi\rangle\) 
can be approximated by linear combinations of states of the form 
\(A(x_1)\dotsm A(x_k)\ket\emptyset\) (any \(k\)). 

By analogy with the Fock construction in the case of free fields, particles 
are usually told to be ``carried'' by fields: the statistics of particles 
(Bose/Fermi) is usually related with the kind of commutation rule at spacelike 
distance. Indeed a much deeper and gratifying 
understanding of statistics as a property
of charged sectors (instead of particle-carrying fields) is available 
(see the review \cite{Doplicher:2009qs}); however 
we will not need such a theoretical deployment here, since we only will be 
concerned with toy models involving the simplest case 
(neutral Klein-Gordon free theory and perturbations).

Unfortunately, the free fields are the only known examples (in dimension 4) 
of theories fulfilling the above general requirements, and they only can 
describe a lifeless universe made of 
non interacting particles. 

A perturbative approach has been developed, where
the interacting dynamics is compared with the free dynamics, as an attempt
to describe scattering experiments: incoming particles which are free 
in the far past and far away from the interaction region 
interact at finite times, and produce 
outgoing particles which are free and far away in the far future. 
The (formal) unitary operator \(S\) 
which interpolates between the incoming and outgoing (asymptotic) 
free states is called the \(S\) matrix. The perturbation series in powers of the
coupling parameter is called the Dyson series.

This perturbative approach also is plagued by terrible problems; 
the formal equations defining the matrix elements of \(S\) are affected by 
all kinds of divergences, and even conceptual obstructions to its existence 
(``Haag theorem'') can be devised. Notwithstanding that, a clever 
strategy (driven by physical interpretation) for covariantly 
removing the most severe class
of divergences (ultraviolet divergences)
has been devised, called renormalisation. A theory is called renormalisable
if the perturbation series for \(S\) can be made to converge, at the cost
of introducing infinite recursive subtractions depending from 
at most a small set of phenomenological parameters. 
Unfortunately the only known renormalisable model for which some indications
about the limit \(S\) can be obtained seems to give \(S=I\), namely
to converge to\ldots a free theory.

Quite surprisingly, however, lowest order corrections in the perturbation 
theory of a physical theory (quantum electrodynamics) 
give experimentally verified predictions of incredible accuracy.

The interested reader will find more details 
on any standard textbook (e.g.\ \cite{Bjorken:1965zz});
the first two chapters of Haag's monograph \cite{Haag:1992hx} 
also provide a nice conceptual
introduction, while the rest of the book is devoted to a mathematically 
advanced introduction to local quantum physics.

\subsection{Why should we quantise the coordinates?}
\label{subsec:why}
Notwithstanding the lack of non trivial well defined models, 
even perturbative, it would be way too much to dismiss the theory as a 
failure. The successful experimental predictions 
should instead be regarded as a strong indication that the illness of the 
theory is due to some conceptual ingredient which is still missing.

The ultraviolet divergences ultimately are a consequence of a too strict
assumption about locality: the divergent expressions arise from interaction
terms which are polynomials in the fields under the pointwise product. String 
theory arose as an attempt to replace points with less singular geometric 
objects. However this solution remains in the realm of classical geometry,
and did not result as effective as it was hoped, so far.

The 
concept itself of space as a collection of infinitely small points dates back 
to Euclid and has never been challenged. Indeed, even Einstein observed
that, as a logical consequence of accepting 
quantum physics,  a quantum description of geometry would be 
conceptually necessary. 
Apart from this, if we take the Compton wavelength \(\lambda_C(m)\)
as the  characteristic parameter driving the quantum behaviour of a point 
particle of  mass \(m\)
and the Schwarzschild radius \(\lambda_C(m)\) 
as characteristic of a pointwise source of gravitational field of mass \(m\), 
the condition \(\lambda_S(m)\sim\lambda_C(m)\) has solution \(m\sim m_P\), the 
Planck mass, and \(\lambda_S(m_P)\sim\lambda_C(m_P)\sim\lambda_P\), where
\(\lambda_P\sim 10^{-33}\text{cm}\) is the Planck length. This is a strong 
indication that quantum phenomena and gravitation come to interplay at the 
Planck scale, where the concept of point particle should be expected to 
break down.

As a first attempt in the direction of concrete physical models, 
it was proposed in \cite{Doplicher:1994tu} to provide a set of non 
commuting coordinates \(q^\mu\) for the flat Minkowski spacetime, 
covariant under a unitary representation of the Poincar\'e group, in order
to replace the usual localisation of quantum fields. 
The hope was to describe an 
intermediate model where the energy involved in single processes is 
sufficiently
high to sense the quantum texture of spacetime; yet the density of processes
is too small to produce curvature. Here we only will describe the solution,
while we refer to the original paper or the less technical \cite{Doplicher:2001qt}
for a derivation of the uncertainty and commutation relations. Earlier attempts
are shortly discussed in \cite{Piacitelli:2010bn}.

Precisely as the components \(x^\mu\)  of the label \(x\) of \(A(x)\) 
are not observables, we are not going to interpret 
the selfadjoint operators \(q^\mu\) as observables. To fix the interpretation,
let us first describe a bit more precisely what happens in usual quantum field
theory.

A localisation state on the classical spacetime may be described by a density 
\(f(x)\,dx\) with \(\int f(x)\,dx=1\), so that the average of the observable
field \(A(x)\) over the density gives the smeared field 
\(A(f)=\int f(x)A(x)\,dx\). A sharp localisation\footnote{We will not concern ourselves here with the 
underlying technicalities (fields should be treated as generalised functions), 
since they are not relevant here.} 
is obtained by a delta: if 
\(\delta_a(x)dx=\delta(x-a)dx\), then \(A(\delta_a)\) is the sharp localisation
of the field \(A(x)\) at \(a\). We are led to think of \(x\) 
as of a set of coordinate functions, of \(f\) as a localisation state,
and of the point \(a\) as of a particular kind of localisation states.

In the case of quantum coordinates \(q^\mu\), localisation states will
be described by density matrices \(\rho\), giving the expectation
\(\langle q^\mu\rangle_\rho=\text{Tr}(\rho q^\mu)\). Vector states are 
a special cases where \(\rho\) is of the form \(|\xi\rangle\langle\xi|\).
To avoid confusion, we anticipate that there will not be a direct 
correspondence between classical sharp localisation states and 
quantum vector states.

Let \(A(x)\) be a quantum field on classical spacetime, taking values in the operators on some Hilbert space \(\mathfrak H\); its DFR quantisation is
the replacement of the classical coordinate functions \(x^\mu\) with the quantum 
coordinate operators \(q^\mu\) on the Hilbert space \(\mathfrak K\), 
using a natural covariant generalisation of the Weyl prescription. If 
\(A(x)=\int\limits_{\mathbb R^4} dk\,\check A(k)e^{ikx}\) , then  
\[
\mathbf A(q)=\int\limits_{\mathbb R^4}dk\,\check A(k)\otimes e^{ikq}
\]
as operators on \(\mathfrak H\otimes\mathfrak K\), where \(kx=k_\mu x^\mu\) and \(kq=k_\mu q^\mu\). States on the second tensor 
factor of \(\mathfrak H\otimes\mathfrak K\) describe a choice of the 
``localisation'' of the field; then for any such choice we get an observable,
whose physical states live in the first tensor factor. In other words,
the expectation functional \(\langle\cdot\rangle_\rho\), acting on the second tensor factor only, plays the same role as the density \(f(x)\,dx\): 
by analogy with the case of classical spacetime, we may introduce the notation
\(\mathbf A(\langle\cdot\rangle_\rho)\) for the partial expectation
\(\langle\mathbf A(q)\rangle_\rho\); then by linearity
\begin{subequations}\label{eq:smeared_qfield}
\begin{equation}
\mathbf A(\langle\cdot\rangle_\rho)=\int\limits_{\mathbb R^4}dk\,\check A(k)\langle e^{ikq}\rangle_\rho=A(f_\rho),
\end{equation}
where 
\begin{equation}
f_\rho(x)=(2\pi)^{-4}\int dk\,e^{-ikx}\langle e^{ikq}\rangle_\rho.
\end{equation}
\end{subequations} 

It seems that the only effect of DFR quantisation is to select a certain subclass of test functions \(f_\rho\) for the usual quantum fields. 
Indeed, the effect of quantisation manifests 
itself when products of fields are considered, as
\[
(\mathbf{AB})(q)\neq \mathbf A(q)\mathbf B(q).
\]
Hence, interaction Lagrangians---which are usually defined in terms of local polynomials
in the fields---have non trivial generalisations to DFR quantum spacetime.

It has to be stressed that, as far as the time component \(x^0\) of classical
localisation in quantum field theory has no interpretation as a time 
observable, so is for its quantum counterpart \(q^0\).

\section{Covariant Quantum Geometry}
\subsection{The DFR quantum coordinates }\label{subsec:dfr_coord}
Consider the operators 
\(P_j=-i\partial_j\), \(Q_j=s_j\cdot\) on 
\(L^2(\mathbb R^4)\), \(j=1,2,3,4\), 
which fulfil \(P_jQ_k-Q_kP_j=-i\delta_{jk}\). We
then introduce the notations \(X^0=P_1,X^1=P_2,X^2=Q_1,X^3=Q_2\). Finally,
we set
\[
\mathfrak K=L^2(\mathscr L,d\varLambda)\otimes 
L^2(\mathbb R^2,ds_1ds_2ds_3ds_4),\]
where \(d\varLambda\) is the Haar measure of the Lorentz group \(\mathscr L\).
As usual, we associate to it a complete set of generalised kets
\[
\ket{\varLambda}\ket{\underline s},\qquad
\varLambda\in\mathscr L,\underline s=(s_1,s_2,s_3,s_4)\in\mathbb R^4,
\]
with bracket
\begin{gather*}
\{\bra{\varLambda}\bra{\underline s}\}\{\ket{\varLambda'}\ket{\underline s'}\}=
\bracket{\varLambda}{\varLambda'}\bracket{\underline s}{\underline s'}
=\delta(\varLambda^{-1}\varLambda')\delta^{(4)}(\underline s_1-\underline s_1'),
\end{gather*}
where integrals are taken with the measure \(d\varLambda d\underline s\) and
\(\delta(\varLambda)d\varLambda\) is the purely atomic normalised
measure on \(\mathscr L\), concentrated on \(I\).

We define the operators \( q^\mu\) by their actions on the kets
\(\ket{\varLambda}\ket{\xi}\):
\begin{equation}\label{eq:def_q}
 q^\mu\ket{\varLambda}\ket{\xi}=\lambda_P\ket{\varLambda}
\{{\varLambda^\mu}_\nu X^\nu\ket\xi\}.
\end{equation}

We may easily check that the operators \( Q^{\mu\nu}\) defined by
\[
[ q^\mu, q^\nu]=i\lambda_P^2 Q^{\mu\nu}
\]
are simultaneously diagonalised by the kets
\(\ket{\varLambda}\ket{\underline s}\):
\[
 Q^{\mu\nu}\ket{\varLambda}\ket{\underline s}=
{\sigma{(\varLambda)}}^{\mu\nu}
\ket{\varLambda}\ket{\underline s},
\]
where 
\begin{subequations}\label{eq:sigma_all}
\begin{equation}\label{eq:sigmanot}
\sigma{(I)}=\begin{pmatrix}0&0&-1&0\\0&0&0&-1\\1&0&0&0\\0&1&0&0\end{pmatrix},\\
\end{equation}
and
\begin{equation}
{\sigma{(\varLambda)}}^{\mu\nu}={\varLambda^{\mu}}_{\mu'}{\varLambda^{\nu}}_{\nu'}
{\sigma{(I)}}^{\mu'\nu'}.
\end{equation}
\end{subequations}
Moreover, by construction
\begin{equation}\label{eq:Q1}
[q^\mu,Q^{\mu\nu}]=0.
\end{equation}
We have a unitary representation \( U(\cdot,0)\) of the Lorentz group
\[
 U(\varLambda,0)\ket M\ket{\underline s}=\ket{\varLambda M}\ket{\underline s};
\]
it fulfils
\begin{gather*}
 U(\varLambda,0)^{-1} q^\mu  U(\varLambda,0)={\varLambda^\mu}_\nu q^\nu,\\
 U(\varLambda,0)^{-1} Q^{\mu\nu} U(\varLambda,0)=
{\varLambda^\mu}_{\mu'}{\varLambda^\mu}_{\mu'} Q^{\mu'\nu'}.
\end{gather*}

Note that, since \({
\sigma{(\varLambda)}}_{\mu\nu}{\sigma{(\varLambda)}}^{\mu\nu}=
{\sigma{(I)}}_{\mu\nu}{\sigma{(I)}}^{\mu\nu}=0\) is a Lorentz invariant, we have
\(Q^{\mu\nu}Q_{\mu\nu}|\varLambda\rangle|\xi\rangle=0\) identically and thus
\begin{subequations}\label{eq:Q}
\begin{equation}\label{eq:Q2}
Q_{\mu\nu}Q^{\mu\nu}=0.
\end{equation}
Analogously,
\begin{equation}\label{eq:Q3}
Q_{\mu\nu}(\ast Q)^{\mu\nu}=\pm 4
\end{equation}
\end{subequations}
where \((\ast Q)^{\mu\nu}=(1/2)\epsilon^{\mu\nu\mu'\nu'}Q_{\mu'\nu'}\).

We finally make use of the remaining Schr\"odinger operators to construct the
representation of translations. We first define
\[
\Pi^0=Q_1+Q_3,\quad\Pi^1=-Q_2-Q_4,\quad\Pi^2=P_1-P_3,
\quad\Pi^3=P_2-P_4,
\]
which fulfil \([\Pi^\mu,\Pi^\nu]=0\) and
\[
[\Pi^\mu,X^\nu]=ig^{\mu\nu}.
\]
Then we define 
\[
p^\mu\{|\varLambda\rangle|\xi\rangle\}=|\varLambda\rangle\{{\varLambda^\mu}_\nu\Pi^\nu|\xi\rangle\}
\]
which fulfil 
\[
[p^\mu,p^\nu]=0,\quad [p^\mu,q^\nu]=ig^{\mu\nu}.
\]
It follows that
\[
U(\varLambda,a)=e^{iap}U(\varLambda,0)
\]
fulfils
\[
U(I,0)=I,\quad U(\varLambda a)U(M,b)=U(\varLambda M, a+\varLambda b)
\]
and
\begin{gather}
 U(\varLambda,a)^{-1} q^\mu  U(\varLambda,a)={\varLambda^\mu}_\nu q^\nu+a^\nu,\\
 U(\varLambda,a)^{-1} Q^{\mu\nu} U(\varLambda,a)=
{\varLambda^\mu}_{\mu'}{\varLambda^\mu}_{\mu'} Q^{\mu'\nu'}.
\end{gather}

Neither the coordinates \(q^\mu\) 
nor the generators of translations \(p^\mu\) 
have an interpretation as observables here. We are not aiming
at a ``more non commutative'' quantum mechanics, but at a noncommutative 
localisation framework for quantum fields.

\subsection{States, uncertainties and optimal localisation}
The operators \(q^\mu\) on \(\mathfrak K\) 
generate all possible localisations  
through the choice of localisation states, namely expectation 
functionals of the form \(\langle\cdot\rangle_\rho=\text{Tr}(\rho\cdot)\) for
some positive operator \(\rho\) with \(\text{Tr}(\rho)=1\) 
(a density matrix). Such a state describes fuzzy localisation around the point
\(a\in\mathbb R^4\) given by \(a^\mu=\langle q^\mu\rangle_\rho\), with variance
\(\Delta_\rho(q^\mu)^2=\langle (q^\mu-\langle q^\mu\rangle_\rho)^2\rangle_\rho\).

By Heisenberg--like arguments, it can be proved that the uncertainties \(\Delta_\rho(q^\mu)\) fulfil the bounds\footnote{There is no reason to expect such bounds to be form--covariant: indeed the uncertainty \(\Delta_\rho\) is not linear 
in its argument, so that \(\Delta_\rho({\varLambda^\mu}_\nu q^\nu)\neq
{\varLambda^\mu}_\nu\Delta_\rho(q^\nu))\) in general.}
\begin{gather}
\Delta_\rho(q^0)(\Delta_\rho(q^1)+\Delta_\rho(q^2)+\Delta_\rho(q^3))\geqslant\lambda_P^2,\\
\Delta_\rho(q^1)\Delta_\rho(q^2)+\Delta_\rho(q^2)\Delta_\rho(q^3)+\Delta_\rho(q^3)\Delta_\rho(q^1)\geqslant\lambda_P^2.
\end{gather}
The proof can be found in \cite{Doplicher:1994tu}.

In the classical case, localisation states arise as probability densities 
\(f(x)dx\) on \(\mathbb R^4\). Points correspond to sharp densities
\(\delta_a(x)=\delta(x-a)dx\). Since \(f\)'s may be rewritten as
\(f(\cdot)=\int da\, f(a)\delta_a(\cdot)\), they can be thought of 
as convex linear  combinations of \(\delta_a\)'s (up to taking limits 
of integral sums). Hence probability densities are the classical analogues of
statistical mixtures and sharp densities are the analogues of  pure states.  

Since however the set of operators \(q^\mu\) is not irreducible 
(by Schur's lemma: the commutators \([q^\mu,q^\nu]\) are not multiples 
of the identity), the usual identification of statistical mixtures with 
non trivial density matrices, and of pure states with vector states, 
breaks down: there are indeed vector states which are not pure!
The pure states are those described by those particular vector states, which 
are of the form
\(|\varLambda\rangle|\xi\rangle\), 
namely they must in particular be eigenkets of the commutators 
\([q^\mu,q^\nu]\).

In view of the large scale limit, 
one would like to have a notion of states with the best possible
localisation properties; then these states could be used to define
interactions with the smallest possible deviations from usual locality.
Indeed, an important condition is that
the usual local quantum theory should appear as a limiting case of the 
new theory (large scale limit).

A moment's though, however, shows that this is impossible, at least in 
such naive terms. Because of the 
uncertainty relations, states are extended objects in some sense, and as such 
they always can be delocalised at wish by suitable Lorentz boosts. 
No such notion as a covariant class
of states with optimal localisation can be devised. This is the fundamental
reason why all more or less 
trivial generalisations of local interactions have led to 
break Lorentz covariance so far. It can be regarded as an indication that the 
noncommutative notion which should give locality in the large scale limit
is non trivial and still missing.

The best one can do is to define well localised states with respect to 
some choice of a class of reference frames which are pairwise related
by a Galilei transformation (no Lorentz boosts). A suitable criterion, then,
is to select those states \(\rho\) 
which minimise \(\sum_\mu\Delta_\rho(q^\mu)^2\). It is clear that these states
are essentially given in terms of the coherent states of the Schr\"odinger
operators \(X^\mu\) used in the construction of the coordinate operators.

Consider first a state of the form \(\ket{I}\ket{\eta_0}\) where
\(\ket{\eta_0}\) is the normalised ground state of the 
harmonic oscillator for the Schr\"odinger operators\footnote{Actually, we should
add a degeneration label \(\kappa\) and write  \(\ket{\eta_0,\kappa}\); since however this degeneracy is only an artefact of the amplification which is used to implement translations, we shall omit the label \(\kappa\).}; then the sum of the squares of the corresponding 
uncertainties attains the minimum 
\[
\sum_\mu\Delta(q^\mu)^2=\lambda_P^2\sum_\mu\Delta(X^\mu)^2=\lambda_P^2\big(\Delta(Q_1)^2+\Delta(Q_2)^2+\Delta(P_1)^2+\Delta(P^2)^2\big)=\lambda_P^2;
\]
a state with these properties is said an optimal localisation state.  
Note that in the states described here 
above, the coordinates have expectation \(0\), so that \(\Delta(q^\mu)\) 
equals the expectation of \({q^\mu}^2\) in such states. 

Next,  we consider the state \(\ket R\ket{\eta_0}\) where 
\(R\in O(3)\subset\mathscr L\). 
For such a state, the coordinates still have expectation in the origin, and
\[
\frac{\{\bra R\bra{\eta_0}\}\sum_\mu (q^\mu)^2\{\ket R\ket{\eta_0}\}}%
{\{\bra R\bra{\eta_0}\}\{\ket R\ket{\eta_0}\}}=
\lambda_P^2\bra{\eta_0}\sum_\mu {R^\mu}_\nu{X^\nu}^2\ket{\eta_0}=
\lambda_P^2\bra{\eta_0}\sum_\mu {X^\mu}^2\ket{\eta_0}=\lambda_P^2;
\]
we used that \(R\) is orthogonal and that \(\sum_\mu {X^\mu}^2=2H_0\) where
\(H_0=\tfrac12(P_1^2+P_2^2+Q^2_1+Q_2^2)\). 
So we still find the ground state of the harmonic oscillator. It follows that
\(\ket R\ket{\eta_a}\) also is an optimal localisation state, with expectation 
in the origin.

Finally for every \(a\in\mathbb R^4\) and any \(R\in O(3)\subset\mathscr L\) 
we define \(|\eta_a\rangle\) by setting
\(\ket R\ket{\eta_a}=U(I,a)\ket R\ket{\eta_0}\); by unitarity, 
\(\ket R\ket{\eta_a}\) is an optimal localisation state, 
but now \(q\) is expected at \(a\).

Indeed, it can be shown (see \cite{Doplicher:1994tu}) that the states described above and their superpositions (with same \(a\)) 
are precisely all the possible optimal localisation states. 
We have
\begin{equation}
\{\langle R|\langle\eta_a|\}e^{ikq}\{|R\rangle|\eta_a\rangle\}=
e^{ika}e^{-\frac1{2}\lambda_P^2\sum_\mu(k^\mu)^2}.
\end{equation}

\subsection{Independent localisation events}\label{subsec:ILE}

The standard way of constructing the coordinates \(q^\mu_j\) of independent events is via 
tensor products, taking
\[
q^\mu_j=I\otimes\dotsm\otimes I\otimes q^\mu\otimes I\otimes\dotsm\otimes I\quad \text{(\(q^\mu\)
in the \(j^\text{th}\) position)},
\]
so that the commutation relations are of the form
\begin{equation}\label{eq:many_Q}
[q^\mu_j,q^\nu_k]=i\delta_{jk}\lambda_P^2Q^{\mu\nu}_j.
\end{equation}

If we take the usual definition of tensor product, we get
\[
Q^{\mu\nu}_j=I\otimes\dotsm\otimes Q^{\mu\nu}\otimes\dotsm\otimes I\quad \text{(\(Q^{\mu\nu}\) in the \(j^\text{th}\) position)}.
\]

However, again 
due to the reducibility of the set \(\{q^\mu\}\) of operators, 
a different construction which also deserves the name of 
tensor product is possible, for which the commutators fulfil
\begin{equation}\label{eq:Qequal}
Q^{\mu\nu}_1=Q^{\mu\nu}_2=\dotsb
\end{equation}
and the \(j\) dependence of \(Q^{\mu\nu}\) in \eqref{eq:many_Q} can be dropped 
\cite{Bahns:2003vb}. 
The idea is to 
define the direct product of kets ``pointwise in \(\varLambda\)'':
\[
\{|\varLambda\rangle|\xi_1\rangle\}\{|\varLambda\rangle|\xi_2\rangle\}\dotsm
\{|\varLambda\rangle|\xi_n\rangle\}=
|\varLambda\rangle|\xi_1\rangle|\xi_2\rangle\dotsm|\xi_n\rangle
\]
These kets span the Hilbert space 
\(L^2(\mathscr L)\otimes L^2(\mathbb R^{4})\otimes\dotsm \otimes
L^2(\mathbb R^{4}) \) (usual direct product).
On such kets, we define the operator 
\[
F_j|\varLambda\rangle|\xi_1\rangle|\xi_2\rangle\dotsm|\xi_n\rangle=
|\varLambda\rangle|\xi_j\rangle|\xi_1\rangle|\xi_2\rangle\dotsm|\xi_{j-1}\rangle|\xi_{j+1}\rangle\dotsm|\xi_n\rangle,
\]
which exchange \(|\xi_1\rangle\) with \(|\xi_j\rangle\).
Finally,
accordingly with the new definition of direct product,
\begin{gather}
q^\mu_1|\varLambda\rangle|\xi_1\rangle|\xi_2\rangle\dotsm|\xi_n\rangle=
\{q^\mu|\varLambda\rangle|\xi_1\rangle\}\rangle|\xi_2\rangle\dotsm|\xi_n\rangle,\\
q^\mu_j=F_jq^\mu F_j,
\end{gather}
which are easily checked to have the desired properties. 

Both choices of \(\otimes\) give covariant coordinates.
In particular for the construction described above---which we will adopt from now on---the representation of the Poincar\'e group is
\[
U(\varLambda,0)|M\rangle|\xi_1\rangle|\xi_2\rangle\dotsm|\xi_n\rangle=
|\varLambda M\rangle|\xi_1\rangle|\xi_2\rangle\dotsm|\xi_n\rangle,
\]
and
\begin{equation}
U(\varLambda,a)^{-1}q^\mu_j U(\varLambda,a)={\varLambda^\mu}_\nu q^\nu_j+a^\mu.
\end{equation}
\subsection{How close can independent events come to?}
\label{subsec:close}
One reason why the direct product ``taken pointwise over \(\varLambda\)'' 
is preferable
when constructing the coordinates of many independent events
is that it leads to a  
natural generalisation of the classical concept of localising independent 
events at the same point \cite{Bahns:2003vb}. 
This can be used for example when a function 
\(f(x_1,x_2,\dotsc,x_n)\) of \(n\) events is evaluated on the diagonal set, 
giving a function \(g(x)=f(x,x,\dotsc,x)\) of one event only.

Define the operator
\[
\bar q^\mu=\frac1{\sqrt n}(q^\mu_1+\dotsb+q^\mu_n).
\]
As a consequence of \eqref{eq:Qequal},
\[
[\bar q^\mu,q^\nu_{j}-q^\nu_k]=0,
\]
namely \(\bar q^\mu\) is statistically independent from the differences of any two events (this would not be the case if the ordinary construction of the direct product were taken instead). 

Note then that every \(q^\mu_j\) can be written as a linear combination
of \(\bar q^\mu\) and the differences \(q^\mu_j-q^\mu_k\):
\[
q^\mu_j=\frac1{\sqrt n}\bar q^\mu+\frac1{n}\sum_k(q^\mu_j-q^\mu_k).
\]

The above remarks suggest to consider a different realisation of the same 
commutation relations, using one more tensor factor.

We define
\begin{equation}\label{eq:tildeq}
\tilde q^\mu_j=\frac1{\sqrt n}q^\mu\otimes\underbrace{I\otimes\dotsm\otimes I}_{\text{\footnotesize \(n\) factors}}\;+\;\frac1{n}\sum_kI\otimes(q^\mu_j-q^\mu_k).
\end{equation}
as operators on \(\underbrace{\mathfrak K\otimes\dotsm\otimes\mathfrak K}_{\text{\footnotesize \(n+1\) factors}}\) (direct products taken ``pointwise in \(\Lambda\)'').
It is clear by construction that
\[
[\tilde q^\mu_j,\tilde q^\nu_k]=i\lambda_P^2\delta_{jk}Q^{\mu\nu},
\]
where \(I\otimes Q^{\mu\nu}\) is identified with \(Q^{\mu\nu}\) according to
\eqref{eq:Qequal}. Moreover the average coordinate is
\[
\frac 1n\sum_j\tilde q_j=\frac1{\sqrt n}q^\mu\otimes\underbrace{I\otimes\dotsm\otimes I}_{\text{\footnotesize \(n\) factors}},
\]
which commutes with the \(\tilde q_j^\mu\)'s. 

The unitary representation
\[
\tilde U(\varLambda,a)=\underbrace{U(\varLambda,a)\otimes\dotsm\otimes U(\varLambda,a)}_{\text{\footnotesize \(n+1\) factors}}
\]
fulfils
\[
\tilde U(\varLambda,a)^{-1}\tilde q^\mu_j\tilde U(\varLambda,a)={\varLambda^\mu}_\nu
\tilde q^\nu_j+a^\mu.
\]

We now want to set all difference \(\tilde q_j-\tilde q_k\) to thei minimum 
value at once, compatibly with the uncertainty relations.

Before giving the general construction, 
we first discuss an easier, less general construction, 
which allows to highlight the  main point. 

We choose  a state \(|R\rangle|\eta_a\rangle\) 
with optimal localisation,
as discussed in section \ref{subsec:ILE}; here \(R\) is an element of \(O(3)\subset\mathscr L\) and \(\eta_a\) is a 
coherent state.

Observe now that the direct product \(\ket\Psi\) 
of \(n\) copies of this state gives,
with our particular definition of direct product, 
\[
\ket\Psi=\underbrace{\{|R\rangle|\eta_a\rangle\}\dotsm\{|R\rangle|\eta_a\rangle\}}_{\text{\footnotesize \(n\) factors}}=
|R\rangle\underbrace{|\eta_a\rangle|\eta_a\rangle\dotsm|\eta_a\rangle}_{\text{\footnotesize \(n\) factors}}.
\]
We can use it to define a partial expectation on the last \(n\) tensor factor.

The components of the separation 
\[
\delta_{jk}\tilde q^\mu=\frac{\tilde q^\mu_j-\tilde q^\mu_k}{\sqrt 2}
\]
between two independent events still fulfil the same relations as the 
coordinates themselves:
\[
[\delta_{jk}\tilde q^\mu,\delta_{jk}\tilde q^\nu]=i\lambda_P^2Q^{\mu\nu}.
\]
Moreover, the partial expectation over the state \(\ket\Psi\) (acting on the last \(n\) tensor factors) gives
\begin{gather*}
\bra\Psi\delta_{jk}\tilde q^\mu\ket\Psi=0,\\
\sum_\mu\Delta(\delta_{jk}\tilde q^\mu)^2=\sum_\mu
\bra\Psi(\delta_{jk}\tilde q^\mu)^2\ket\Psi=\lambda_P^2,
\end{gather*}
and the latter is precisely the property selecting the states with optimal localisation;
note that, as expected, the choice of \(a\) is irrelevant.
Hence we may say that this partial expectation has the effect of setting
the differences \(\delta\tilde q_{jk}\) as close to zero as possible, compatibly with the uncertainty relations. We regard this as a quantum generalisation
of the classical operation of setting \(x_1=x_2=\dotsc\). 

This is almost what 
we want; the only problem is that there is no need to restrict
to a particular joint eigenspace of the \(Q^{\mu\nu}\)'s, namely
the one corresponding to the 
projection on \(\ket R\). It is sufficient to restrict to the sum of all 
joint eigenspaces of the \(Q^{\mu\nu}\)'s which correspond to orthogonal 
transformations. 

To do this, we split the above operation in two steps. We first define
the orthogonal projection \(E\) which 
sends
\(|\varLambda\rangle|\xi_1\rangle|\xi_2\rangle\dotsm|\xi_n\rangle\) to 0
if \(\varLambda\) contains a Lorentz boost, and leaves it unchanged otherwise.
We have
\begin{gather*}
[\tilde q^\mu_j,E],\\
[\tilde U(R,a),E]=0,\quad R\in O(3),a\in\mathbb R^4.
\end{gather*}

In other words, \(E\) is the biggest possible projection which commutes with
all \(q_j^\mu\)'s and is stable under orthogonal transformations. 
We then restrict our 
coordinates \(\tilde q^\mu_j\) to operators acting on 
the range \(L^2(O(3))\otimes L^2(\mathbb R^{4(n+1)})\) of \(E\), 
and afterwards we  take the partial expectation on the state
\(\ket{\eta_a}\dotsm\ket{\eta_a}\) (\(n\) factors) acting on the last \(n\) direct factors. The resulting map \(\mathbb E\) has its range in the operators
on \(L^2(O(3))\otimes L^2(\mathbb R^4)\), and 
has the following properties:
\begin{gather}
\mathbb E[\delta_{jk}\tilde q^\mu]=0,\\
\label{eq:ultra_gauss}
\mathbb E[e^{ik_\mu\delta_{jk}\tilde q^\mu}]=e^{-\lambda_P^2\frac12\sum_\mu (k^\mu)^2}
\end{gather}
as multiples of the identity operator. 

Moreover
\[
\mathbb E\left[\frac{\tilde q^\mu_1+\dots+\tilde q^\mu_n}n\right]\ket R\ket\xi=
\frac{\lambda_P}{\sqrt n}\ket R\{{R^\mu}_\nu X^\nu\ket\xi\},
\]
which defines new operators 
\begin{equation}
\dot q^\mu=\mathbb E\left[\frac{q_1+\dotsb+q_n}n\right]
\end{equation}
on \(L^2(O(3))\otimes L^2(\mathbb R^4)\) with the nice property that the
corresponding commutators
\[
[\dot q^\mu,\dot q^\nu]=i\left(\frac{\lambda_P}{\sqrt n}\right)^2\dot Q^{\mu\nu}
\] 
induce the same uncertainty relations of the initial coordinates, but with
the Planck length scaled by \(\sqrt n\). This is precisely what one would expect
of the statistical behaviour of a mean of independent stochastic variables. 
In the large \(n\) limit, the average coordinate of many events becomes
deterministic.

Note that the coordinates \(\dot q^\mu\) are covariant under orthogonal 
transformations and translations:
\begin{subequations}
\begin{equation}
\dot U(R,a)^{-1}\dot q^\mu\dot U(R,a)={R^\mu}_\nu\dot q^\nu+a^\mu,\quad R\in O(3)\subset\mathscr L, a\in\mathbb R^4,
\end{equation}
where the representation \(\dot U(R,a)\) is obtained by restricting each unitary
operator \(U(R,a)\) to \(L^2(O(3))\otimes L^2(\mathbb R^4)\).

Finally, the map \(\mathbb E\) is covariant in the sense that
\begin{equation}
\dot U(R,a)^{-1}\mathbb E[\,\cdot\,]\dot U(R,a)=
\mathbb E[\tilde U(R,a)^{-1}\,\cdot\, \tilde U(R,a)],\quad R\in O(3)\subset \mathscr L,a\in\mathbb R^4.
\end{equation}
\end{subequations}

We will use this map to define a quantum generalisation of the Wick product
in section \ref{subsec:quantum_wick}, where we will need the explicit form
of \(\mathbb E[e^{i\sum_jk_jq_j}]\) which we will now compute.

We first map \(e^{i\sum_j k_jq_j}\) (which acts on kets of the form
\(\ket\varLambda\ket{\xi_1}\dotsm\ket{\xi_n}\)), into \(e^{i\sum_jk_j\tilde q_j}\)
(which acts on kets of the form
\(\ket\varLambda\ket\xi\ket{\xi_1}\dotsm\ket{\xi_n}\)).

We observe that 
\[
\sum_jk_j\tilde q_j=\left(\frac{1}{\sqrt n}\sum_jk_j\right) q\otimes
\underbrace{I\otimes\dotsm\otimes I}_{\text{\(n\) factors}}\;+\;
I\otimes\sum_j\left(k_j-\frac 1n\sum_l k_l)q_l\right). 
\]
acting on \(\ket R\ket{\xi}\ket{\xi_1}\dotsm\ket{\xi_n}\). It follows that
\[
e^{i\sum_jk_j\tilde q_j}=\left(e^{\frac i{\sqrt n}\big(\sum_jk_j\big) q}\right)\otimes
\left(e^{i\sum_j\left(k_j-\frac 1n\sum_l k_l\right)q_l}\right). 
\]
where the direct product is taken ``pontwise in \(\varLambda\)''.

Now we restrict to the range of the projection \(E\). We do that 
simply by  restricting ourselves from now on 
to kets \(\ket{R}\) with \(R\in O(3)\subset \mathscr L\).

We  take the 
partial expectation of \(e^{i\sum_jk_j\tilde q_j}\) (restricted to the range of \(E\)) over the last \(n\) factors,
using a state \(\ket{\eta_0}\dotsm\ket{\eta_0}\) where \(\eta_0\) is the 
ground state of the harmonic oscillator: this gives 
\begin{align}
\nonumber\mathbb E[e^{i\sum_jk_jq_j}]&=e^{i\sum_jk_j\dot q}
\{\underbrace{\bra{\eta_0}\dotsm\bra{\eta_0}}_{\text{\footnotesize{\(n\) factors}}}\}\sum_j\left(k_j-\frac 1n\sum_l k_l\right)q_l
\{\underbrace{\ket{\eta_0}\dotsm\ket{\eta_0}}_{\text{\footnotesize{\(n\) factors}}}\}=\\ &=
e^{-\lambda_P^2\frac12\left(\sum_j|k_j|^2-\sum_{jl}k_j\cdot k_l\right)}e^{i\sum_jk_j\dot q}
\label{eq:E_weyl}
\end{align}
where \(h\cdot k=\sum_{\mu=0}^3k^\mu h^\mu\), \(|k|^2=\sqrt{k\cdot k}\), and
we recall that \(\dot q^\mu\) is the restriction of \(q^\mu/\sqrt n\) to 
\(L^2(O(3))\otimes L^2(\mathbb R^4)\).

\subsection{Distance, area and volume operators}
In the framework of the universal differential calculus of \cite{dubois-violette:univ-calc}, we may define the differential of coordinates as
\begin{equation}
dq^\mu=q^\mu\otimes I-I\otimes q^\mu.
\end{equation}
This provides another reason why the construction of \(\otimes\) which we used
in section \ref{subsec:ILE} is preferable: if it is used in the 
definition of \(d\), then \(dQ^{\mu\nu}=0\), which is compatible with the 
interpretation of \(Q^{\mu\nu}\) as an independent background.

If the universal calculus is used alone, the realisation of the commutation
relations by operators on the Hilbert space plays little role. 
An interesting way of making the differential calculus 
to interplay with operator products is to use the operator product instead
of the tensor product when multiplying differentials with each other \cite{4-volume}, e.g.
\begin{align*}
dq^\mu\cdot dq^\nu&=(q^\mu\otimes I-I\otimes q^\mu)\cdot
(q^\nu\otimes I-I\otimes q^\nu)=\\
&=q^\mu\otimes q^\nu\otimes I-q^\mu\otimes I\otimes q^\nu-I\otimes q^\mu q^\nu\otimes I+I\otimes q^\mu\otimes q^\nu.
\end{align*}
So the product of two differentials is a combination of products of 
operator living on the 3-fold tensor product of the one--event state space; 
which is consistent with the interpretation of \(dq\) as a ``segment'' with two extreme events,
the product of two differentials describing the ``join'' of 
two such ``segments'' at the same event.

In particular, a very simple generalisation of the usual definitions of 
area and 3- and 4-volume  operators can be given \cite{4-volume}. For example,
the 4-volume operator is defined as
\begin{equation}
V=\epsilon_{\mu\nu\rho\sigma}dq^\mu\cdot dq^\nu\cdot dq^\rho\cdot dq^\sigma,
\end{equation}
which lives in the 5-fold tensor product, and indeed one needs five events
to give a hypercube in four dimensions.

The resulting operator is not selfadjoint, as a consequence of the commutators which show up when exchanging the order of the ``vertexes''. Quite unexpectedly,
\(V\) is normal, namely \(VV^*=V^*V\). The phase operator 
appearing in the polar decomposition of the 4-volume operator can be regarded
as a quantum generalisation of the sign describing the orientation.

The 4-volume operator is very complicated; yet its spectrum can be computed and 
is found to be 
\(\{(n\sqrt 5\pm 2+ia)\lambda_P^4:n\in\mathbb Z,a\in\mathbb R\}\). In particular
the absolute value of the 4-volume operator is bounded below by \((\sqrt 5-2)\lambda_P^4\approx .23 \lambda_P^4\).
We refer to the original paper \cite{4-volume} for the details of the computation. 

\subsection{The $\star$-product}\label{subsec:star}
Let 
\begin{equation}\label{eq:def_Sigma}
\Sigma=\{\sigma{(\varLambda)}:\varLambda\in\mathscr L\},
\end{equation}
according to the notation \eqref{eq:sigma_all}.

Provided that the integrals exist, we may associate to each
complex function \(f(\sigma;x)\) of 
\(\Sigma\times\mathbb R^4\) 
the operator 
\begin{align}\nonumber
f(Q;q)\{\ket\varLambda\ket\xi\}&=
\ket\varLambda\left\{\int dk{}\check f(\sigma{(\varLambda)};k)
e^{i\lambda_P(\varLambda X)k}\ket\xi\right\}=\\
&=\ket\varLambda\left\{\int dk{}\check f(\sigma{(\varLambda)};k)
e^{i\lambda_PX(\varLambda^{-1}k)}\ket\xi\right\},
\end{align}
where the Fourier transformation acts on \(f(\sigma;\cdot)\) 
for every \(\sigma\) fixed.
Note that, if \(f\) does not depend explicitly on \(\sigma\), the above
is simply \(\int dk\,{}\check f(k)e^{ikq}\); if, otherwise, \(f\) 
does not depend explicitly on \(x\), the above is the usual 
function \(f(Q)\) of the sixteen pairwise commuting operators \(Q^{\mu\nu}\).

Now we consider the product of two such operators
\begin{align}\nonumber
f(Q;q)&g(Q;q)\{\ket\varLambda\ket\xi\}=\\
&=\ket\varLambda\left\{\int dh\,dk{}
\check f(\sigma{(\varLambda)};h)
\check g(\sigma{(\varLambda)};k)
e^{i\lambda_PX(\varLambda^{-1}h)}e^{i\lambda_PX(\varLambda^{-1}k)}\ket\xi\right\}
\label{eq:prod_op}
\end{align}
By definition the operators \(X^\mu\) (defined at the beginning of section
\ref{subsec:dfr_coord}) fulfil the commutation relations \([X^\mu,X^\nu]=
{\sigma(I)}^{\mu\nu}\), where \(\sigma{(I)}\) is given by \eqref{eq:sigmanot}.
The BCH formula implies
\[
e^{i\lambda_PhX}e^{i\lambda_PkX}=e^{-\frac i2\lambda_P^2
\sigma{(I)}^{\mu\nu}h_\nu k_\nu}e^{i(h+k)X};
\]
substituting this in \eqref{eq:prod_op}, we get
\begin{equation}
f(Q;q)g(Q;q)\{\ket\varLambda\ket\xi\}=
\ket\varLambda\left\{\int dh\,dk{}(\check f\tilde\times\check g)(\sigma{(\varLambda)};k)
e^{i\lambda_PkX)}\ket\xi\right\},
\end{equation}
where
\begin{equation}
(\check f\tilde\times\check g)(\sigma;k)=
\int dh \check f(\lambda_P^2\sigma;h)\check g(\lambda_P^2\sigma;k-h)
e^{-\frac i2\lambda_P^2\sigma^{\mu\nu}h_\nu k_\nu},
\end{equation}
and antisymmetry of \(\sigma\) has been used. 

Defining now
\begin{equation}
(f\star g)(\sigma;x)=\int dk(\check f\tilde\times\check g)(\sigma;k)e^{ikx},
\end{equation}
we get
\begin{equation}
f(Q;q)g(Q;q)=(f\star g)(Q;q).
\end{equation}
Note that, even in the case when  \(f\) and \(g\) do not explicitly depend 
on \(\sigma\), their  \(\star\)-product does. 

For the explicit expression of \(\star\) and its (complicate) relationship
with the Moyal expansion, see \cite{Piacitelli:2010bn}.
The \(\star\)-product here only plays an ancillary role. Its only use in this
paper is the following. To every \(f(Q;q)\) and \(t\in\mathbb R\) we can
associate the operator
\[
\int\limits_{x^0=t} d^3x\,f(Q;x)
\]
on \(L^2(\mathscr L)\)
whose action is
\[
\int\limits_{x^0=t} d^3x\,{}f(Q;x)\ket\varLambda=
\int\limits_{x^0=t} d^3x\,{}f(\sigma{(\varLambda)};x)\ket\varLambda. 
\]
This map is positive, in the sense that it maps \(f(Q;q)^*f(Q;q)\) 
(which is a positive operator) to another positive operator (see \cite[Sec. 5]{Doplicher:1994tu} for 
the proof). 
We only will need the 
\(\star\)-product in the case of operators of the form
\(f_1(Q;q)\dotsm f_n(Q;q)=(f_1\star\dotsm \star f_n)(Q;q)\), which the above
map sends into
\(\int_{x^0=t}(f_1\star\dotsm \star f_n)(Q;x)\).

\section{Quantum Field Theory on Quantum Spacetime}

\subsection{The Klein--Gordon field on classical spacetime}
To fix the notations, we briefly recall the standard definition of the massive
free scalar spin 0 field (see any standard textbook, e.g.~\cite{Bjorken:1965zz}).

We establish a Dirac bracket notation which defines the (Fock) Hilbert 
space~\(\mathfrak H\). The complete system of kets 
\begin{align*}
&\ket\emptyset,&n=0,\\
&|\vec k_1,\dotsc,\vec k_n\rangle,\quad &n=1,2,\dotsc,\quad \vec k_j\in\mathbb R^3
\end{align*}
fulfils the normalisation condition
\begin{eqnarray}\nonumber
\lefteqn{\langle \vec h_1,\vec h_2\dotsc,\vec h_m|\vec k_1,\vec k_2,\dotsc,\vec k_n\rangle=}\\
&=&
\frac{\delta_{nm}}{n!}\sum_\pi\delta^{(3)}(\vec h_1-\vec k_{\pi_1})\delta^{(3)}(\vec h_2-\vec k_{\pi_2})\dotsm\delta^{(3)}(\vec h_n-\vec k_{\pi_n}),
\end{eqnarray}
where the sum runs over all permutations \(\pi\) of \((1,\dotsc,n)\), and the 
\(\delta^{(3)}\)'s are defined with respect to the usual translation invariant 
measure \(d^3k\) on \(\mathbb R^3\). In particular for any such \(\pi\) we have \(|\vec k_1,\dotsc,\vec k_n\rangle=
|\vec k_{\pi_1},\dotsc,\vec k_{\pi_n}\rangle\).
The ket \(|\emptyset \rangle\) is called the vacuum state. 

The creation and annihilation operators
\begin{subequations}
\begin{align}
a^\dagger(\vec k)|\vec k_1,\dotsc,\vec k_n\rangle&=\sqrt{n+1}|\vec k,\vec k_1,\dotsc,\vec k_n\rangle,\\
a(\vec k)|\vec k_1,\dotsc,\vec k_n\rangle&=\sqrt{\frac1n}\sum_j\delta^{(3)}(k-k_j)|\vec k_1,\dotsc,\vec k_{j-1},\vec k_{j+1},\dotsc,\vec k_n\rangle
\end{align}
\end{subequations}
fulfil 
\begin{subequations}
\begin{gather}a^\dagger(k)=a(k)^*,\\
[a(\vec h),a^\dagger(\vec k)]=\delta^{(3)}(\vec h-\vec k),\\
\quad [a(\vec h),a(\vec k)]=
[a^\dagger(\vec h),a^\dagger(\vec k)]=0,\\
a(\vec k)|\emptyset\rangle=0.
\end{gather}\end{subequations}
The scalar Klein-Gordon field is defined as
\begin{subequations}
\begin{equation}
\varphi(x)=\int\limits_{\mathbb R^4} d^4k\,\check\varphi(k)e^{ikx}
\end{equation}
where
\begin{equation}
\check\varphi(k^0,\vec k)=\sqrt{\frac{2|k^0|}{(2\pi)^3}}\delta(k^2-m^2)\{\theta(-k^0)a(-\vec k)+\theta(k^0)a^\dagger(\vec k)\}.
\end{equation}
\end{subequations}
Note that to each 3-vector \(\vec k\) there is a unique 4-vector \(\tilde k\)
belonging to the upper mass shell (namely fulfilling \(k^0>0\) and \(k^2={k^0}^2-|\vec k|^2=m^2\)). Explicitly, \(\tilde k=(\omega_m(\vec k),\vec k)\), where
\(\omega_m(\vec k)=\sqrt{m^2+|\vec k|^2}\).

Since \(\square e^{ikx}=-k^2e^{ikx}\) (where \(\square=\partial_\mu\partial^\mu\)), we have
\begin{equation}
(\square+m^2)\varphi=0
\end{equation}

There is a representation \(\mathscr U(\varLambda,a)\)
of the restricted Poincar\'e group \(\mathscr P_+^\uparrow\) (where only
Lorentz transformations preserving the arrow of time 
and the orientation of space are allowed) by unitary operators on \(\mathfrak H\): its action is
\begin{subequations}
\begin{gather}
\mathscr U(\varLambda,a)\ket \emptyset =\ket \emptyset ,\\
\mathscr U(\varLambda,a)\ket{\vec k_1,\dotsc,\vec k_n}=e^{i(\sum_j\varLambda\tilde k_j)a}\ket{
\overrightarrow{\varLambda \tilde k_1},
\overrightarrow{\varLambda \tilde k_2},\dotsc,\overrightarrow{\varLambda \tilde k_n}}.
\end{gather}
\end{subequations}
The notation is a bit involved, let us describe what happens: we start from a 3-vector \(\vec k\), we construct the corresponding on-shell 4-vector \(\tilde k\), we apply to it the Lorentz 
transformation \(\varLambda\) giving the 4-vector \(\varLambda\tilde k\), whose space part is 
\(\overrightarrow{\varLambda\tilde k\phantom{_j}}\). Note that, since \(\varLambda\) preserves the time arrow, \(\varLambda\tilde k\) is in the upper 
mass shell, too.

By construction,
\[
\mathscr U(\varLambda,a)a^\dagger(\vec k)\mathscr U(\varLambda,a)^{-1}=
e^{i(\varLambda\tilde k)a}a^\dagger\Big(\overrightarrow{\varLambda\tilde k\,}\Big).
\]
It follows that\footnote{To check these computations the matrix notation for Lorentz matrices can be useful: the Lorentzian product \(ka=k_\mu a^\mu\) may be written as \(K^t GA\), where \(K,A\) are column vectors, \(t\) is matrix transposition, 
and \(G=\text{diag}(1,-1,-1,-1)\) is the metric. Then conservation of the metric is \(\varLambda G\varLambda^t=G\). From this and \(G^2=I\) follows 
\(\varLambda^{-1}=G\varLambda^t G\), which in turn gives (back to usual notations) \(k(\varLambda a)=
(\varLambda^{-1}k)a\).}
\begin{subequations}
\begin{align}
\mathscr U(\varLambda,a)\check\varphi(k)\mathscr U(\varLambda,a)^{-1}
&=e^{i(\varLambda k)a}\check\varphi(\varLambda k),\\
\mathscr U(\varLambda,a)\varphi(x)\mathscr U(\varLambda,a)^{-1}&=
\varphi(\varLambda x+a)),\\
\end{align}
\end{subequations}
where \((\varLambda,a)\in\mathscr P_+^\uparrow\).

The spectrum condition holds, namely the generators \(P^\mu\) of translations 
defined by \(\mathscr U(I,a)=e^{iaP}\) fulfil \(P_\mu P^\mu\geqslant 0\) and \(P^0\geqslant 0\). In particular we have \(P^\mu\ket\emptyset=0\) and
\begin{equation}
P^\mu\ket{\vec k_1,\dotsc,\vec k_n}=\sum_{j}{\tilde k_j}^\mu
\ket{\vec k_1,\dotsc,\vec k_n};
\end{equation}
note that any finite sum \(\sum_j{\tilde k_j}^\mu\) is contained in 
the convex hull of the upper mass shell of mass \(m\).

It is noteworthy for our purposes that the Hamiltonian \(H_0=P^0\)
of the free field takes the form
\begin{subequations}
\begin{equation}
H_0=\int\limits_{x^0=t}d^3x\;{}\mathscr H_0(x),
\end{equation}
where 
\begin{equation}
\mathscr H_0(x)=\frac12\int\limits_{x^0=0}d^3x\wick{(\partial^0\varphi)(x)^2-(\partial^0\partial^0\varphi)(x)\varphi(x)}
\end{equation}
\end{subequations}
and the double dots indicate normal (Wick) ordering of annihilations and 
creations. 
\subsection{DFR quantisation of the Klein--Gordon field}

According to the discussion of the introduction,
\begin{equation}
\boldsymbol\varphi(q)=\int dk\,\check\varphi(k)\otimes e^{ikq}
\end{equation}
as an operator on \(\mathfrak H\otimes\mathfrak K\), where \(\mathfrak H\) 
is the Fock space and the coordinates \(q^\mu\) are operators on 
\(\mathfrak K\). In particular, for \(\ket{k_1,\dotsc,k_n}\in\mathfrak H\)
and \(\ket\varLambda\ket\xi\in\mathfrak K\),
\begin{equation}
\boldsymbol\varphi(q)\{|k_1,\dots,k_n\rangle|\varLambda\rangle|\xi\rangle\}=
\int dk\{\check\varphi(k)|k_1,\dotsc,k_n\rangle\}\{e^{ikq}|\varLambda\rangle|\xi\rangle\}.
\end{equation}

By abuse of notations, we still denote by \(\mathscr U,U\)
 the representations of Poincar\'e transformations acting non trivially on 
the first and second tensor factor of \(\mathfrak H\otimes\mathfrak K\), 
respectively.

We find
\begin{align*}
\mathscr U(\varLambda,a)\boldsymbol\varphi(q)
\mathscr U(\varLambda,a)^{-1}&=\int dk\,\left(\mathscr U(\varLambda,a)
\check\varphi(k)\mathscr U(\varLambda,a)^{-1}\right)\otimes e^{ikq}=\\
&=\int dk\,e^{i(\varLambda k)a}\check\varphi(\varLambda k)\otimes e^{ikq}=\\
&=\int dh\,\check\varphi(h)\otimes e^{ih(\varLambda q+a)}=\\
&=\int dh\,\check\varphi(h)\otimes 
\left(U(\varLambda,a)^{-1}e^{ihq}U(\varLambda,a)\right)=\\
&=U(\varLambda,a)^{-1}\boldsymbol\varphi(q)U(\varLambda,a).
\end{align*}
Hence the DFR quantisation of a free quantum field is covariant. Note
that the above result can be given the simpler form
\begin{equation}
\mathscr U(\varLambda,a)\boldsymbol\varphi(q)
\mathscr U(\varLambda,a)^{-1}=\boldsymbol\varphi(\varLambda q+a).
\end{equation}

Since translations are unitarily implemented, we also have derivatives; we find
that the DFR quantisation commutes with taking derivatives:
\begin{equation}
\lim_{\lambda\rightarrow 0}\frac1\lambda(\mathbf\varphi(q+\lambda e^\mu)= 
(\boldsymbol{\partial_\mu\varphi})(q),
\end{equation}
where \(e^\mu\) is the \(\mu^{\text{th}}\) canonical basis vector of \(\mathbb R^4\).

Hence, the DFR quantised free field fulfils the Klein--Gordon equation:
\begin{equation}
(\square + m^2)\boldsymbol\varphi(q)=0.
\end{equation}

The partial expectation of \(\boldsymbol\varphi(q)\) over a state with optimal localisation around \(a\in\mathbb R^4\) gives, according to \eqref{eq:smeared_qfield}, the free field operator on the Fock space \(\mathfrak H\), smeared with
a Gaussian:
\begin{align}\nonumber
F_a=\{ \bra R \bra{\eta_a} \} \boldsymbol\varphi(q)\{\ket R\ket{\eta_a}\}
&=\int dk\,{}
\check\varphi(k)
\{\bra R\bra{\eta_a}\}
e^{ikq}\{\ket R\ket{\eta_a}\}=\\
&=\frac1{(2\pi)^{2}}\int dx\; 
\varphi(x)e^{-\frac{|x-a|^2}{2\lambda_P^2}},
\end{align}
where \(|x-a|^2=\sum_\mu(x^\mu-a^\mu)^2\). It follows that
\begin{equation}
[F_a,F_{a+x}]\propto i\frac{\lambda_P}{|\vec x|}\left(e^{-\frac1{8\lambda_P^2}(|\vec x|-x^0)^2}-e^{-\frac1{8\lambda_P^2}(|\vec x|+x^0)^2}\right)
\end{equation}
which falls off as a Gaussian in spacelike dimensions (as a function of \(x\)),
and converges to the usual commutator function in the large scale limit.

\subsection{The DFR perturbative setup}\label{subsec:dfr_setup}

The basic idea of the DFR perturbative setup is to construct an {\itshape 
effective non 
local quantum field theory on the classical spacetime}: the underlying idea is
the following: there are incoming and outgoing free fields on the classical spacetime, describing free particles: since they do not interact, they 
``do not know'' that the spacetime is quantum. When interaction takes place,
the quantum texture of spacetime enters in the game, and this is taken into 
account by a nonlocal deformation of the interaction Lagrangian.

A key remark is that
\begin{equation}
\boldsymbol H_0(Q)=\int\limits_{x^0=t}d^3x\mathscr H_0(q)=H_0\otimes I,
\end{equation}
namely the free Hamiltonian is left essentially unchanged by the Weyl quantisation, apart from the
\(\otimes I\) which reminds that, while \(H_0\) is an operator on the Fock space \(\mathfrak H\), its quantisation is an operator on \(\mathfrak H\otimes\mathfrak L^2(\mathscr L)\). 

Since \(Q^{\mu\nu}\) is unaffected by time translations, we have
\begin{equation}
e^{it\boldsymbol H_0(Q)}\boldsymbol\varphi(q^0,q^1,q^2,q^3)e^{-it\boldsymbol H_0(Q)}=
\boldsymbol\varphi(q^0+t,q^1,q^2,q^3).
\end{equation}
Free fields are essentially unaffected by the DFR quantisation.

This remark led the authors of \cite{Doplicher:1994tu} to continue 
this analogy, and define
a nonlocal generalisation of the interaction Lagrangian 
\(\wick{\varphi(x)^n}\) as 
\(\wick{\boldsymbol\varphi(q)^n}\), which can be written equivalently
as \(\wick{(\varphi\star\dotsm\star\varphi}(Q;q)\). 
Then, by the remarks at the end 
of section \ref{subsec:star}, it makes sense to consider 
\begin{equation}
\boldsymbol H_{I,\star}(Q;t)=\int\limits_{x^0=t}d^3x
\wick{(\varphi\star\dotsm\star\varphi)}(Q,x),
\end{equation}
again as a (formal) operator on \(\mathfrak H\otimes\mathfrak L^2(\varLambda)\), and we add the subscript \(\star\) to keep track of the choice of the noncommutative Wick product. 

At a certain point, we will have 
to take a partial expectation so to obtain an effective
interaction term \(H_{I,\star}^{\text{eff}}(t)\) as an operator on the Fock 
space \(\mathfrak H\); in other words, we have to integrate out the \(Q\) 
dependence. 
This is necessary to obtain a scattering matrix which interpolates 
the incoming and outgoing free fields, which live on the Fock space
\(\mathfrak H\) alone; 
indeed, physical intuition suggests that the noncommutativity
gets averaged out over large distances without interactions. However, let us 
leave this apart for a while.

Mimicking usual QFT, we formally define the limit 
\(\boldsymbol S_{\star}(Q)=\boldsymbol U_{\star}(Q;\infty,-\infty)\) 
of the unitary evolution semigroup \(\boldsymbol U_{\star}(Q;t,s)\), 
which fulfils
\begin{gather*}
\boldsymbol U_{\star}(Q;t,t)=\boldsymbol I,\\
\boldsymbol U_{\star}(Q;s,t)\boldsymbol U_{\star}(Q;t,u)=\boldsymbol U_{\star}(Q;s,u),\\
\boldsymbol U_{\star}(Q;s,t)^{-1}=\boldsymbol U_{\star}(Q;t,s),
\end{gather*} 
and solves the evolution equation
\begin{equation}
\frac{\partial \boldsymbol U_{\star}}{\partial t}(Q;t,s)=i\boldsymbol H_{I,\star}(Q;t)\boldsymbol U_{\star}(Q;t,s)
\end{equation}
Its formal solution is given by the Dyson series 
\begin{equation}
\boldsymbol S_{\star}(Q)=I+\sum_{N=1}^\infty\frac{i^N}{N!}\int dt_1\dotsm dt_N\,T[\boldsymbol H_{I,\star}(t_1),\dotsc,\boldsymbol  H_{I,\star}(t_N)],
\end{equation}
where \(T\) means that the product of the \(\boldsymbol H_{I,\star}(t_j)\)'s is taken in the order
of decreasing times, namely
\begin{eqnarray}\nonumber
\lefteqn{T[\boldsymbol H_{I,\star}(t_1),\dotsc,\boldsymbol  H_{I,\star}(t_N)]=}\\
&=&
\sum_\pi\theta(t_{\pi_1}-t_{\pi_2})\theta(t_{\pi_2}-t_{\pi_3})\dotsm \theta(t_{\pi_{n-1}}-t_{\pi_n})\boldsymbol H_{I,\star}(t_{\pi_1})\boldsymbol H_{I,\star}(t_{\pi_2})\dotsm \boldsymbol H_{I,\star}(t_{\pi_n})\nonumber\\
&&
\end{eqnarray}
where \(\theta\) is Heaviside's step function and the sum runs over the permutations of \((1,\dotsc,n)\).

Note that the time ordered product is taken with respect to the labels
\(t_j\) at which the space integral is taken. 

The matrix elements
of \(\boldsymbol S_{\star}\) (as an operator on the Fock space, 
times \(L^2(\mathscr L)\)) are of the form
\begin{equation}
\{\bra{h_1,\dotsc,h_m}\bra\varLambda\}\boldsymbol S_{\star}\{\ket{k_1,\dotsc,k_n}\ket M\}=\delta(\varLambda^{-1}M)\bra{h_1,\dotsc,h_m}S_{\star}(\sigma(\varLambda))
\ket{k_1,\dotsc,k_n}
\end{equation}
where \(S_{\star}(\sigma)\) is a non local, non causal 
scattering matrix on the usual Fock space. Note that the dependence on \(\lambda_P\) is hidden in the product \(\star\). 
Indeed, for every \(\sigma\in\Sigma\) we have the formal limit 
\begin{equation}
S_{\star}(\sigma)\underset{\lambda_P\rightarrow 0}{\longrightarrow} S_{\text{loc}},
\end{equation}
where \(S_{\text{loc}},\) is the local, causal scattering  matrix
of the (non renormalised) \(\varphi^n_4\) theory on the classical Minkowski 
spacetime.
 
The situation is much alike that of a bundle of non local theories over 
\(\Sigma\). To integrate out this dependence, one would like to take a partial
expectation on the second tensor factor with some Lorentz 
invariant state in \(L^2(\mathscr L)\). 
Unfortunately, no such state exists, essentially because 
the Lorentz group is not amenable\footnote{This means that there is no 
left-invariant mean of the functions of \(\mathscr L\). 
Of course there is the Haar measure \(d\varLambda\), but a mean should send, by definition, the constant function 1 to 1, while 
\(\int d\varLambda\, 1=\infty\).}. The most symmetric choice is to take the 
rotation invariant state described by the characteristic function of 
the set \(O(3)\subset\mathscr L\), which
is square summable since \(O(3)\) is compact. The partial
expectation of \(\boldsymbol S_{\star}\) on such state 
defines an effective scattering matrix 
\begin{equation}
S_{\star}^{\text{eff}}=\int\limits_{\Sigma^{(1)}} d\sigma\,S_{\star}(\sigma)
\end{equation}
on the Fock space \(\mathfrak H\), where \(\Sigma^{(1)}=
\{\sigma(R):R\in O(3)\subset\mathscr L\}\) and \(d\sigma\) is the invariant measure on \(\Sigma\) induced by the Haar measure of \(\mathscr L\), normalised so that \(\Sigma^{(1)}\) has measure 1. This gives a theory covariant 
under rotations, but not under Lorentz 
boosts\footnote{Some authors claim to obtain a fully covariant interacting 
theory by means of a Lorentz invariant measure \(W(\theta)d\theta\),
where \(d\theta\) is the invariant measure on the space \(\mathcal T\) of
real second rank antisymmetric
tensors. Unfortunately, such a Lorentz invariant measure must be of the form
\(W(\theta)d\theta=w(\theta^{\mu\nu}\theta_{\mu\nu},(\theta^{\mu\nu}(\ast\theta)_{\mu\nu})^2)d\theta\) for some \(w(a,b)\). Hence 
\[
\int\limits_{\mathcal T} W(\theta)d\theta\;f(\theta)=
\int\limits_{-\infty}^{\infty}da\int\limits_0^\infty db\,w(a,b)
\int\limits_{\mathcal T_{a,b}}d\theta_{a,b}f(\theta),
\]
where
\(\mathcal T_{a,b}=\{\theta:\theta^{\mu\nu}\theta_{\mu\nu}=a,(\theta^{\mu\nu}(\ast\theta)_{\mu\nu})^2=b)\}\) and \(d\theta_{a,b}\) is the measure induced by \(d\theta\)
on \(\mathcal T_{a,b}\). For every \(a,b\), \(\int_{\mathcal T_{a,b}}d\theta_{a,b}=\infty\), hence this measure, though Lorentz invariant, does not define a mean.
The situation is similar to the absence of a translation invariant normalised
measure on the line: there's no such thing as the 
expected position of a particle equally distributed on the line.}.

If we perform a shameless exchange \(\int d\sigma\sum_N=\sum_N\int d\sigma\)
in the Dyson series,
the above would be the same as starting from the beginning with a non local
effective interaction term
\begin{equation}
H_{I,\star}^{\text{eff}}(t)=\int\limits_{\Sigma^{(1)}}d\sigma\;{}
\int\limits_{x_0=0} d^3x\,
\wick{\varphi\star\dotsm\star\varphi}(\sigma;x)
\end{equation}
on the Fock space \(\mathfrak H\). 

Since all these developments are formal, there is room for experimentation 
about the order of summation and integrations. Choosing to integrate out
the \(\sigma\) dependence separately at each vertex after having performed 
Wick  reduction, the effective scattering matrix was found ultraviolet finite
in the \(\varphi_4^3\) theory \cite{Bahns:2004zb}.

It is clear that, even if the effective theory is regular, its large scale 
limit will reproduce the non renormalised theory. Hence finite renormalisation 
would anyway be necessary, where the finite subtractions should diverge 
in the large scale limit so to reproduce the infinite subtractions of usual
renormalisation.

\subsection{Unitarity and Feynman diagrams}
Since the Hamiltonian is formally selfadjoint, the scattering matrix
is formally unitary, and no violations of unitarity should be expected. 
The violations discussed in the literature  
may be regarded as a consequence of an inconsistent prescription 
for the time ordering (as pointed out in \cite{Bahns:2002vm}).

Indeed, for a local theory, the second order contribution to the Dyson series 
is
\[
S_{\text{loc},2}=\frac12\iint ds\,dt\,T[H_I^{\text{loc}}(s),H_I^{\text{loc}}(t)].
\]
Since the time ordering does nothing to a pointwise product of \(n\) fields, 
namely \(T[\varphi(x),\dotsc,\varphi(x)]=\varphi(x)^n\), the above can be written
as 
\begin{align*}
S_{\text{loc},2}&=\frac12\iint ds\,dt\,\int\limits_{x^0=s} d^3x\int\limits_{y^0=t} d^3y\;{}\;{}
T[\wick{\varphi(x)^n}\,,\,\wick{\varphi(y)^n}]=\\
&=\frac12\int d^4x\int d^4y\;{}\;{}
T[\wick{\varphi(x)^n}\,,\,\wick{\varphi(y)^n}],
\end{align*}
where the time ordering is with respect to the times \(x^0,y^0\).
In other words, the time ordering can be brought inside integrals.

For the  non local theory described by the interaction term \(H_{I,\star}^{\text{eff}}(t)\), this cannot be done, since the latter is of the form
\begin{equation}
H_{I,\star}^{\text{eff}}=\int da_1\dotsm da_n\;w(a_1,\dotsc,a_n;t)
\wick{\varphi(a_1)\dotsm \varphi(a_n)}
\end{equation}
for some totally symmetric kernel \(w\), 
depending on the time parameter \(t\) (and implicitly on \(\lambda_P\)). 
The time ordering
is taken with respect to such time parameters, not to the time components 
\(a_j^0\) of the integration variables.

Indeed, using the proper integral form of \(\star\), there's no such object as
a ``\(T[\varphi(x_1)\star\dotsm\star\varphi(x_n)]\)''. 
But if instead one {\itshape illegally} uses the Moyal 
expansion of \(\star\), then he/her is misled to think that the noncommutative
product is defined pointwise, since the twist only contains derivatives; in 
which case one would find ``\(T[\varphi(x)\star\dotsm\star\varphi(x)]=
\varphi(x)\star\dotsm\star\varphi(x)\)''
(false) and again could safely bring the time ordering inside the integral (false). This is one possible mechanism to obtain the violations of unitarity. 
For a general discussion of the drawbacks of the Moyal expansion, see 
\cite{Piacitelli:2010bn}.

Upon inserting the Dyson series in the Gell-Mann\&Low formula, the usual 
diagrammatic expansion may be used, with minor modifications to the rules. 
To every vertex, associate a factor
\(w(a_1,\dotsc,a_n;t)da_1\dotsm da_n\);
to every line originating from that vertex and labeled by \(x\) pick a factor
\(\frac1i(\Delta_+(x-a_j)\theta(x^0-t)+\Delta_+(a_j-x)\theta(t-x^0)\); between any line connecting the vertex with another vertex \(w(b_1,\dotsc,b_n;s)db_1\dotsm db_n\), pick a factor \(\frac1i(\Delta_+(b_k-a_l)\theta(s-t)+\Delta_+(a_l-n_k)\theta(t-s)\); and for any two external lines labeled by \(x,y\),
pick a usual Stueckelberg-Feynman propagator \(\Delta_{SF}(a_j-b_k)\). See 
\cite{Piacitelli:2004rm}
for a detailed discussion.

\subsection{Quantum Wick Products and ultraviolet regularity}
\label{subsec:quantum_wick}
It is peculiar of the process of generalisation that 
equivalent procedures may have inequivalent
generalisations. Consider \(n\) functions \(f_j(x_j)\) of independent variables,
and define \(F(x_1,\dotsc,x_n)=f_1(x_1)\dotsm f_n(x_n)\). The evaluation
of \(F\) at coincident points  may be described either as 1)  
setting \(x_1=x_2=\dotsc=x_n=x\), or 2) evaluate the pointwise 
product of functions \(f_1\dotsm f_n\) at 
\(x\). The equivalence of these two procedures is summarised by
\[
f_1(x)\dotsm f_n(x)=(f_1\dotsm f_n)(x),
\]
which is so natural that on first sight we do not even notice the point.

The non commutative generalisation of the ``product strategy'' is to replace
the pointwise product with the product of quantised
functions, or equivalently with the star product, so that one obtains
\[
(f_1\star\dotsm\star f_n)(Q;q).
\]
This was used in the definition of non local Wick product
\(\wick{\varphi\star\dotsm\star\varphi}\) discussed in the
preceding section. 

A non commutative generalisation of the 
``bring independent events to the same place'' strategy instead may be 
given in terms of the map \(\mathbb E\) discussed in section 
\ref{subsec:close}, and gives a different result. It was used in \cite{piac_tesi,Bahns:2003vb} to 
obtain a different generalisation of Wick product. 

The idea is to define the quantum wick product as
\begin{equation}
\qwick{\varphi(\dot q)^n}=
\mathbb E[\wick{\boldsymbol\varphi(q_1)\dotsm\boldsymbol\varphi(q_n)}],
\end{equation}
where it is understood that \(\mathbb E\) acts on the localisation part.

Since
\begin{align*}
\wick{\boldsymbol\varphi(q_1)\dotsm\boldsymbol\varphi(q_n)}&=
\int d^4k_1\dotsm d^4k_n\wick{\check\varphi(k_1)\dotsm\check\varphi(k_n)}\otimes
e^{ik_1q_1}\dotsm e^{ik_nq_n}=\\
&=\int d^4k_1\dotsm d^4k_n\wick{\check\varphi(k_1)\dotsm\check\varphi(k_n)}\otimes
e^{i\sum_jk_jq_j},
\end{align*}
we get by linearity and \eqref{eq:E_weyl}
\begin{align*}
\mathbb E[\wick{\boldsymbol\varphi(q_1)\dotsm\boldsymbol\varphi(q_n)}]
&=\int d^4k_1\dotsm d^4k_n\wick{\check\varphi(k_1)\dotsm\check\varphi(k_n)}\otimes
\mathbb E[e^{i\sum_jk_jq_j}]=\\
&=\int d^4k_1\dotsm d^4k_n\;e^{-\lambda_P^2\frac12\left(\sum_j|k_j|^2-\sum_{jl}k_j\cdot k_l\right)}
\wick{\check\varphi(k_1)\dotsm\check\varphi(k_n)}\otimes e^{i(\sum_jk_j)\dot q}.
\end{align*}
Standard Fourier theory now gives
\[
\mathbb E[\wick{\boldsymbol\varphi(q_1)\dotsm\boldsymbol\varphi(q_n)}]=
\qwick{\underbrace{\varphi\star\dotsm\star\varphi}_{\text{\footnotesize \(n\) factors}}}(\dot q)
\]
where 
\begin{eqnarray}\nonumber
\lefteqn{\qwick{\underbrace{\varphi\star\dotsm\star\varphi}_{\text{\footnotesize \(n\) factors}}}(x)=}\\\nonumber
&=&\frac{n^2}{(2\pi)^{8(n-1)}}\int da_1\dotsm da_n\,\wick{\varphi(x+a_1)\dotsm\varphi(x+a_n)}
e^{-\frac1{2\lambda_P^2}\sum_j|a_j|^2}\delta^{(4)}\left(\frac1{n\lambda_P}\sum_ja_j\right)\\
&&
\end{eqnarray}
With this definition, we obtain 
an interaction term of the form
\begin{equation}
\boldsymbol H_{I,\mathscr G}(t)=H_{I,\mathscr G}^{\text{eff}}(t)\otimes \dot I
\end{equation}
as an operator on \(\mathfrak H\otimes L^2(O(3))\), where
\begin{eqnarray}\nonumber
\lefteqn{H_{I,\mathscr G}^{\text{eff}}(t)=\int\limits_{x^0=0} d^3x\,\qwick{\varphi\star\dotsm\star\varphi}(x)=}\\
&=&\nonumber
\frac{n^2}{(2\pi)^{8(n-1)}}\int\limits_{x^0=t}d^3x\int da_1\dotsm da_n
\wick{\varphi(x+a_1)\dotsm
  \varphi(x+a_n)}\\
&&e^{-\frac1{2\lambda_P^2}\sum_j|a_j|^2}\delta^{(4)}
\left(\frac1{n\lambda_P}\sum_ja_j\right).
\end{eqnarray}

Here covariance under Lorentz boosts is broken by the map \(\mathbb E\).
The resulting effective interaction term \(\boldsymbol H_{I,\mathscr G}^{\text{eff}}(t)\) does not depend
explicitly on \(\sigma\in\Sigma^{(1)}\). Hence the average over the invariant 
measure gives precisely \(H_{I,\mathscr G}^{\text{eff}}(t)\) which defines
a non local, non causal scattering matrix on the Fock space \(\mathfrak H\).

Note that the only effect of noncommutativity here is to naturally reproduce
a particular recipe for the so called ``point-split regularisation''
in terms of a \(\mathscr G\)aussian kernel.
The resulting \(H_{I,\mathscr G}^{\text{eff}}(t)\) is completely free of ultraviolet
divergences. However, the same remarks apply here about the need of finite 
renormalisation, as in the end of section \ref{subsec:dfr_setup}.

\footnotesize
\bibliographystyle{utphys}
\bibliography{mybibtex}
\end{document}